\documentclass[journal=jacsat,manuscript=article]{achemso}
\usepackage[version=3]{mhchem} 
\usepackage{amsmath}
\usepackage{bm}
\usepackage{enumerate}
\usepackage{graphicx}
\usepackage{color}

\usepackage{cleveref}
\usepackage{mathrsfs}
\usepackage{xcolor}
\usepackage{natbib}
\usepackage[utf8]{inputenc}

\author{Daisy Q. Wang}
\affiliation{School of Physics, University of New South Wales, Sydney, NSW 2052, Australia}
\alsoaffiliation{Australian Research Council Centre of Excellence in Future Low-Energy Electronics Technologies, University of New South Wales, Sydney 2052, Australia}
\author{Zeb Krix}
\affiliation{School of Physics, University of New South Wales, Sydney, NSW 2052, Australia}
\alsoaffiliation{Australian Research Council Centre of Excellence in Future Low-Energy Electronics Technologies, University of New South Wales, Sydney 2052, Australia}
\author{Oleg P. Sushkov}
\affiliation{School of Physics, University of New South Wales, Sydney, NSW 2052, Australia}
\alsoaffiliation{Australian Research Council Centre of Excellence in Future Low-Energy Electronics Technologies, University of New South Wales, Sydney 2052, Australia}

\author{Ian Farrer}
\altaffiliation{Present Address: Department of Electronic and Electrical Engineering, The University of Sheffield, Mappin Street, Sheffield, S1 3JD, United Kingdom}  

\author{David A. Ritchie}
\affiliation{Cavendish Laboratory, J. J. Thomson Avenue, Cambridge, CB3 0HE, United Kingdom}

\author{Alexander R. Hamilton}
\affiliation{School of Physics, University of New South Wales, Sydney, NSW 2052, Australia}
\alsoaffiliation{Australian Research Council Centre of Excellence in Future Low-Energy Electronics Technologies, University of New South Wales, Sydney 2052, Australia}

\author{Oleh Klochan}
\email{o.klochan@unsw.edu.au}
\affiliation{School of Science, University of New South Wales, Canberra ACT 2612, Australia}
\alsoaffiliation{Australian Research Council Centre of Excellence in Future Low-Energy Electronics Technologies, University of New South Wales, Sydney 2052, Australia}

\title{Formation of artificial Fermi surfaces with a triangular superlattice on a conventional two dimensional electron gas}

\begin{document}

\begin{abstract}

In nearly free electron theory the imposition of a periodic electrostatic potential on free electrons creates the bandstructure of a material, determined by the crystal lattice spacing and geometry. Imposing an artificially designed potential to the electrons confined in a GaAs quantum well makes it possible to engineer synthetic two-dimensional band structures, with electronic properties  different from those in the host semiconductor. Here we report the fabrication and study of a tuneable triangular artificial lattice on a GaAs/AlGaAs heterostructure where it is possible to transform from the original GaAs bandstructure and Fermi surface to a new bandstructure with multiple artificial Fermi surfaces simply by altering a gate bias. For weak electrostatic potential modulation magnetotransport measurements reveal quantum oscillations from the GaAs two-dimensional Fermi surface, and classical oscillations due to these electrons scattering from the artificial lattice. Increasing the strength of the modulation reveals new quantum oscillations due to the formation of multiple artiﬁcial Fermi surfaces, and ultimately to new classical oscillations of the electrons from the artificial Fermi surface scattering from the superlattice modulation.  These results show that low disorder gate-tuneable lateral superlattices can be used to form artiﬁcial two dimensional crystals with designer electronic properties.
\end{abstract}


The field of 2D materials has received a significant boost with the advent of Moir\'{e} superlattices~\cite{Dean13}. Stacking two atomically thin materials on top of each other with a twist angle between them creates an additional long range potential modulation – the Moir\'{e} superlattice. The twist angle between the 2D layers defines the period of the Moir\'{e} superlattice, modifying the energy bands and hence the optoelectronic properties. A spectacular example is the case of two single layer graphene sheets stacked on top of each other with a twist angle of 1.08 degrees~\cite{Cao18}, called magic angle twisted bilayer graphene. The electronic properties of this system can be tuned from a correlated insulator into a superconductor just by changing a gate voltage~\cite{Cao18_2}. The origin of this extraordinary behaviour is strong electron-electron interactions due to a flat band formed in the energy spectrum induced by the Moir\'{e} superlattice at this very specific twist angle~\cite{Bistritzer11}. Since the original discovery, other “twisted” 2D material systems have been shown to form flat bands with a variety of correlated phases, making twisted 2D systems the ideal playground to study these exotic electronic phases~\cite{Liu20, Wang20_2}. However, Moir\'{e} superlattices have some limitations: 1) the symmetry of the Moir\'{e} superlattice is set by the constituent 2D materials; 2) the strength of the periodic superlattice potential is fixed by the interaction between the two 2D layers and cannot be varied; 3) the Moire superlattice requires accurate control of the twist angle between the two layers~\cite{Cao18, Cao18_2, Lu19}, but very often the fabrication process results in a spatially varying twist angle and thus a spatially non-uniform superlattice~\cite{Uri20}. 

One approach to overcome the limitations of twisted Moir\'{e} systems is to use nanolithographically patterned lateral superlattices~\cite{Forsythe18, Jessen19, Huber22}. This technique allows complete control over the lattice symmetry, lattice constant, and strength of the 2D periodic lattice potential. It can also be easily integrated into conventional semiconductor heterostructures, which have the advantage of well-established fabrication technology and superb device quality. Early studies of one-dimensional lateral superlattices in semiconductor heterostructures resulted in the observation of commensurability oscillations in the longitudinal resistance, when the cyclotron orbit of electrons matches the lattice spacing of the artificial 1D lattice ~\cite{Gerhardts89, Winkler89}, and culminated with signatures of Hofstadter physics in the longitudinal and Hall resistances of 2D lateral superlattices ~\cite{Albrecht99, Deutschmann01, Geisler04}. However, limitations of materials and fabrication technologies, such as disorder in the patterned superlattice and the weak strength of the lateral superlattice potential, prevented the study of bandstructures in 2D artifical crystals.

The difficulty of creating artificial bandstructures in such a system lies in fabricating an extremely uniform and small-period superlattice while maintaining low disorder(high electron mobility). Both disorder in the patterned superlattice and in the heterostructure can impede the observation of artificial bands. Besides nanofabrication requirements for superlattice patterning (the superlattice period $a$ should be $\sim 100 \textrm{nm}$ or less), modulation-doped heterostructures should also be avoided to suppress long-range charge disorder due to the existence of randomly positioned ionised dopants~\cite{Ashwin20}. Furthermore, the energy gap created by a periodic modulation potential is directly related to the strength of the potential, which decays exponentially as a function of the distance between the superlattice and the 2D electron system~\cite{Tkachenko15}. Therefore, extremely shallow heterostuctures ($d \ll a$) are essential to achieve a strong enough modulation potential to experimentally observe artificial bandstructure.

Here we overcome these challenges by combining a recently developed ultra-shallow high-mobility 2D electron gas in undoped GaAs/AlGaAs heterostructures with high resolution electron beam lithography, to create an artificial two-dimensional crystal with a gate-tunable 2D lateral superlattice. 

\begin{figure}[h!]
\centering
\includegraphics[width=\textwidth]{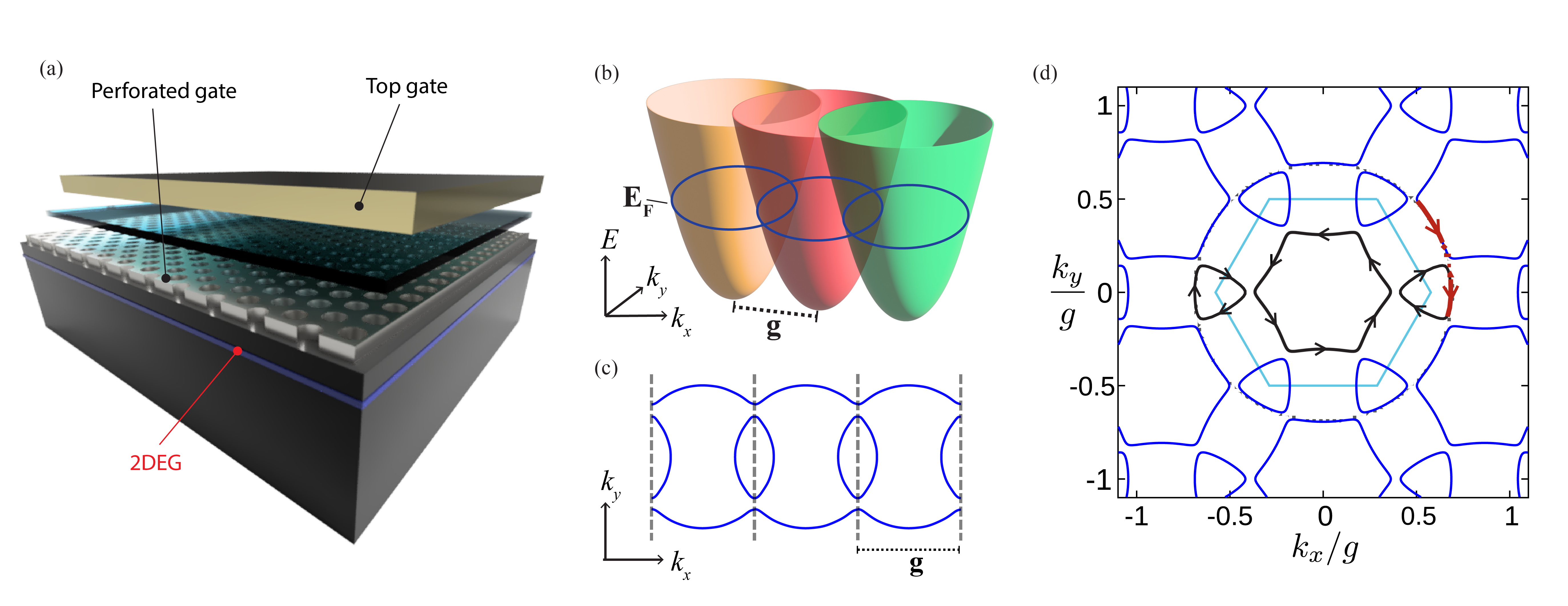}
\caption{(a) Schematic of the device layout. The device has a double-layer design utilising a bottom perforated gate (PG) and an unstructured top gate (TG). The two gates are separated by a thin layer of $\mathrm{AlO_x}$. 
(b) Energy bands of an electron system with one-dimensional periodic modulation. Fermi surface indicated in blue. (c) Restructuring of Fermi surface topology for a one-dimensional modulation as the periodic potential strength is increased. Energy gaps in the spectrum open where the rings intersect. (d) Reconstructed Fermi surfaces of a 2D triangular lattice. Black lines indicate closed orbits involving no magnetic breakdown. Red arrows illustrate the process of magnetic breakdown in the presence of a weak magnetic field. Due to magnetic breakdown, electrons can tunnel through the energy gap which leads to more complicated electron orbits.}
\label{fig:device}
\end{figure} 

A typical device structure in our study is shown in Figure~\ref{fig:device}(a). The shallow GaAs/AlGaAs accumulation-mode heterostructure is completely undoped and consists of a GaAs buffer, $\mathrm{30\ nm}$ of AlGaAs and a $\mathrm{7\ nm}$ GaAs cap. The 2DEG has a mobility of $1.5 \times10^{6}\ \mathrm{cm^2/Vs}$ and a mean free path of $11\ \mu \mathrm{m}$ at a density of $2\times10^{11}\ \mathrm{cm}^{-2}$. A thin layer of Ti is deposited on top of the heterostructure and patterned with a triangular array of holes with a lattice constant of $80\ \mathrm{nm}$ and a hole diameter of  $30\ \mathrm{nm}$. This perforated gate is used to induce carriers at the hetero-interface and tune the carrier concentration. A $15\ \mathrm{nm}$ $\mathrm{AlO_x}$ dielectric is then grown on the Ti gate, followed by an overall Ti/Au top gate.  The top gate is used to vary the strength of the modulation potential since it is fully screened by the perforated Ti gate except where the holes are. This double-layer design allows both the carrier density and the modulation strength to be varied separately~\cite{Wang20}. 

When a periodic modulation potential is applied to the 2DEG, the parabolic energy bands of free electrons are shifted by a multiple of the reciprocal lattice vector $\bm{g} = (2\pi /a, 0)$, where $a$ is the lattice constant of the periodic modulation. Therefore, the Fermi surface reconstructs into a set of intersecting rings of radius $k_F$, plotted in blue in Figure~\ref{fig:device}(b) for a one-dimensional lattice. When the strength of the potential modulation is increased, energy gaps open where these rings intersect, and the topology of the Fermi surface changes into a closed elliptical Fermi surface and open Fermi surface as shown in Figure~\ref{fig:device}(c). The restructured Fermi surface for a two dimensional triangular lattice is shown in Figure~\ref{fig:device}(d). Here, the circular shape of the original Fermi surface is evident, but the actual Fermi surface is made up of non-circular closed orbits (black lines in Figure~\ref{fig:device}(d)), which are separated by energy gaps. In the presence of a weak magnetic field, more complicated trajectories can form when electrons undergo quantum tunnelling from one orbit to another (Figure~\ref{fig:device}(d)), a process know as magnetic breakdown~\cite{Steda90}.  Any closed trajectories obtained from sections of the original circular Fermi surface linked by quantum tunnelling become a new reconstructed Fermi surface which can quantise and form discrete energy levels. These levels cause quantum (Shubnikov-de Haas) oscillations of the magnetoresistance in an unpatterned 2DEG with a frequency proportional to the area of the orbit and are commonly used to measure the area of the circular Fermi surface. Therefore, the existence of reconstructed Fermi surfaces can also be detected via quantum oscillations of different frequencies.

\begin{figure}[h!]
\centering
\includegraphics[width=\textwidth]{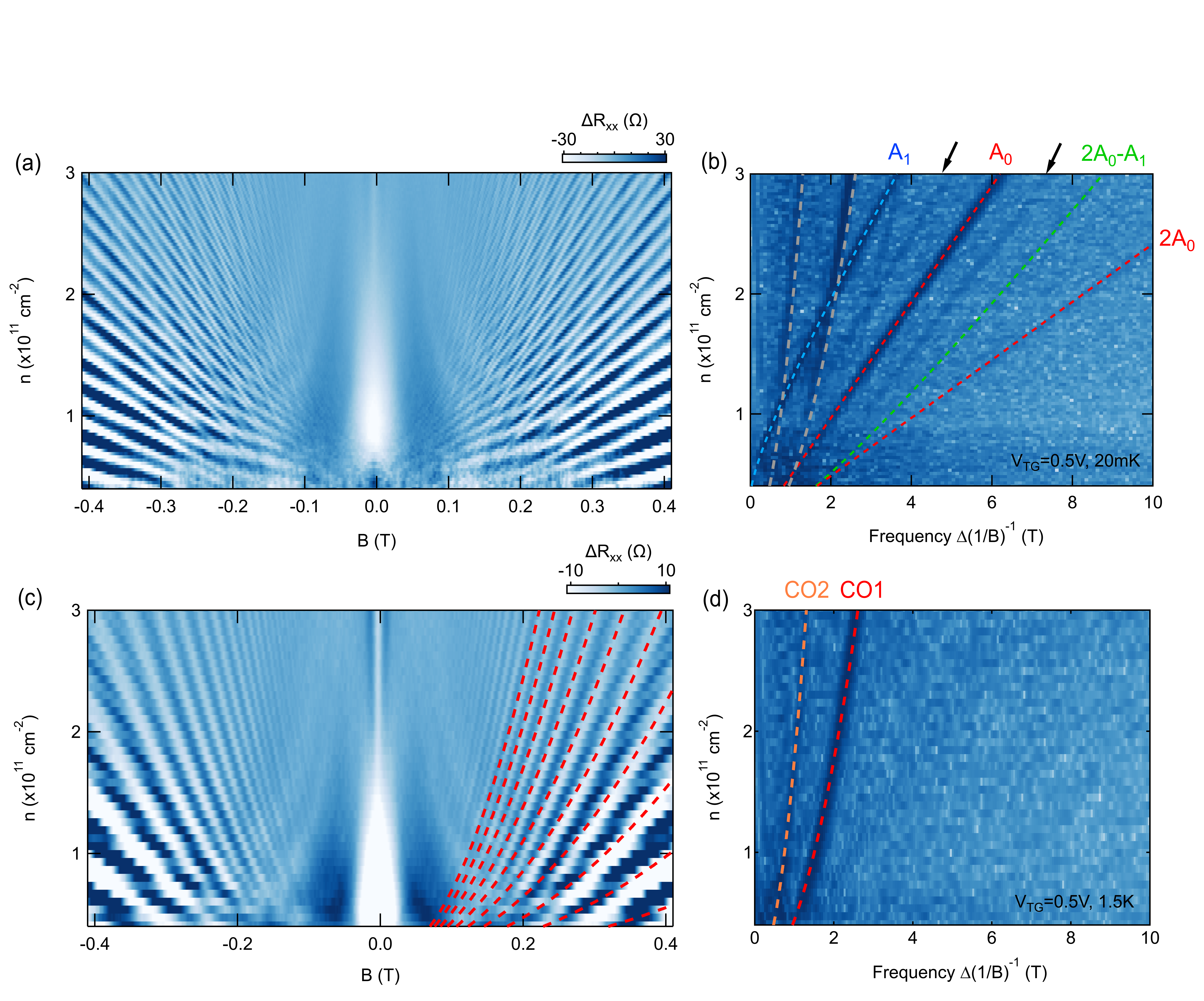}
\caption{(a) Background removed magneto-resistance $\Delta R_{xx}$ at $20\ \mathrm{mK}$, as a function of the carrier density, $n$, and perpendicular magnetic field, $B$ for $\mathrm{V_{TG}=0.5\ V}$. (b)  Fourier transform of the magnetoresistance in $1/B$, as a function of carrier density and frequency $\Delta(\frac{1}{B})^{-1}$. Grey dashed lines are the theoretically calculated frequencies of commensurability oscillations of the original Fermi surface $A_0$. Red, blue and green dashed lines are theoretically calculated frequencies of quantum oscillations from different Fermi surfaces $A_0$, $A_1$ and $2A_0-A_1$ respectively. Black arrows indicate the position of half frequencies. (c) Background removed magneto-resistance $\Delta R_{xx}$ measured at $1.5\ \mathrm{K}$. Dashed lines indicate the predicted locations of minima due to commensurability oscillations using Eq.~(\ref{eq:co1}). (d) Fourier transform of the magnetoresistance at $1.5\ \mathrm{K}$. The red (orange) dashed line is the calculated frequency of commensurability oscillations for single (double) lattice planes.}
\label{fig:weak}
\end{figure}

Figure~\ref{fig:weak}(a) shows the experimentally measured magnetoresistance of the device, as a function of the carrier density, $n$, and the perpendicular magnetic field, $B$, at $20\ \mathrm{mK}$. A smoothed background is removed from $R_{xx}$ to make the magnetoresistance oscillations in $\Delta R_{xx}$ more visible (See supplementary~\ref{fig:supp background} for details). In Figure~\ref{fig:weak}, $\mathrm{V_{TG}}$ is set to $0.5\ \mathrm{V}$ which corresponds to a weak modulation potential. At this modulation strength, $\Delta R_{xx}$ shows a complicated braiding pattern signalling the existence of oscillations at different frequencies. To extract the individual frequency components the Fourier transform of $\Delta R_{xx}(1/B)$ is computed. 
The amplitude of the frequency components is plotted in logarithmic colour scale in Figure~\ref{fig:weak}(b), where large amplitude appears as dark blue. As shown in Figure~\ref{fig:weak}(b), multiple frequency components can be clearly identified over a wide density range.

\begin{figure}[h!]
\centering
\includegraphics[width=\textwidth]{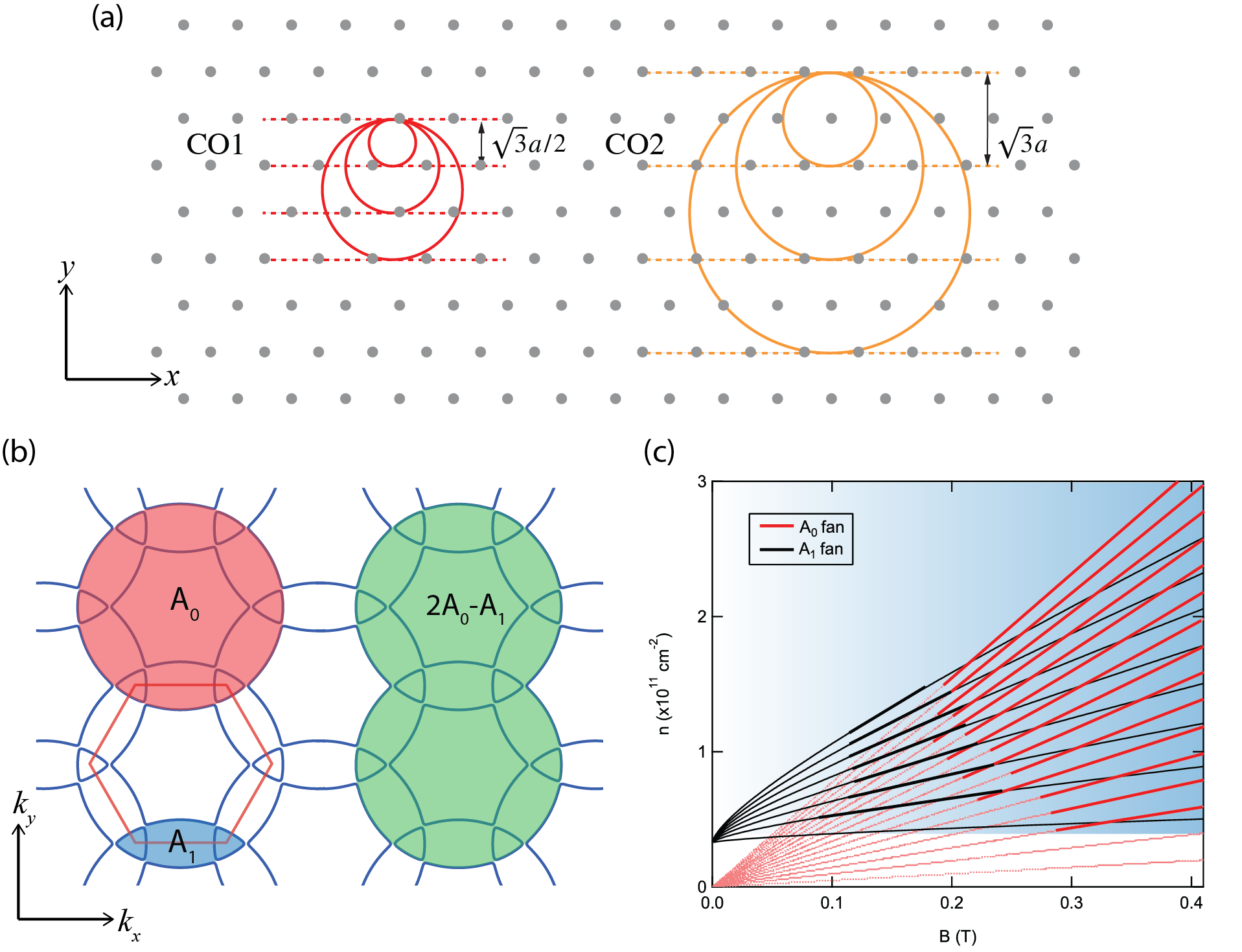}
\caption{(a) A schematic showing the two commensurate conditions for a 2D triangular lattice: red for the commensurate condition between any consecutive lattice planes ($L=\sqrt{3}/2a$) and orange for the commensurate condition where every second lattice plane is considered ($L=\sqrt{3}a$) due to the symmetry of a triangular lattice. (b) A schematic in the reciprocal space showing the original circular Fermi surface $A_0$ and reconstructed Fermi surfaces $A_1$ and $2A_0-A_1$. (c) Calculated fan diagram resulting from two different Fermi surfaces $A_0$ and $A_1$ highlighting maxima of $\Delta R_{xx}$ observed in Figure~\ref{fig:weak}(a). The features associated with the $A_0$ ($A_1$) fan are coloured in red (black). Blue-shaded area corresponds to the experimentally accessible density range in Figure~\ref{fig:weak}.}
\label{fig:Fermi}
\end{figure}

To understand the origin of these different frequency components, the same measurement is repeated at a higher temperature of $1.5\ \mathrm{K}$. At $1.5\ \mathrm{K}$, all low-field quantum oscillations are suppressed due to thermal excitation, leaving only classical oscillations. Figures~\ref{fig:weak}(c) and (d) plot $\Delta R_{xx}$ and the Fourier transform of $\Delta R_{xx}(1/B)$ at $1.5\ \mathrm{K}$ respectively. The two frequency components in Figure~\ref{fig:weak}(d) which also exist at $20\ \mathrm{mK}$ are highlighted by the grey dashed lines in Figure~\ref{fig:weak}(b). These two frequency components are caused by classical commensurability oscillations. In a semi-classical picture, the drift velocity perpendicular to the lattice plane vanishes when the cyclotron orbit is commensurate with the distance between lattice planes (Figure~\ref{fig:Fermi}(a)), resulting in resistance minima~\cite{Beenakker89}. This commensurability occurs when the cyclotron radius $R_{c}$ is related to the spacing between lattice planes $L$ by
\begin{equation}
    2R_c = (\lambda - 1/4)L
\label{eq:co1}
\end{equation}
where $\lambda$ is an integer.
Substituting $R_c=\hbar k_F/eB$, it can be easily shown that commensurability oscillations are periodic in $\mathrm{1/B}$ with a periodicity of
\begin{equation}
\Delta \left( \frac{1}{B} \right) =\frac{eL}{2\hbar k_F}
\label{eq:co2}
\end{equation}
which are the grey dashed lines plotted in Figure~\ref{fig:weak}(b).

For a 2D triangular superlattice with lattice constant $a$, the spacing between the fundamental set of lattice planes is $L = \sqrt{3}/2a$ (Figure~\ref{fig:Fermi}(a)). The frequency calculated from Eq.(\ref{eq:co2}), CO1, is plotted as the red dashed line in Figure~\ref{fig:weak}(d) and is in excellent agreement with the experiment with no fitting parameters. The locations of resistance minima are also very well predicted by Eq.~(\ref{eq:co1}), shown as the red dashed lines in Figure~\ref{fig:weak}(c). The other frequency component highlighted by the orange dashed line in Figure~\ref{fig:weak}(d), CO2, corresponds to a new commensurate condition that is unique to the 2D triangular superlattice. Due to the symmetry of the triangular lattice, consecutive lattice planes are not equivalent. From one plane to the next there is a horizontal shift of the lattice sites by $a/2$. This feature of the the 2D triangular superlattice gives rise to another set of lattice planes with spacing $L' = 2 L$. Thus, two sets of commensurability oscillations are observed in the experiment with one having exactly half the frequency of the other. 

Excluding these commensurability oscillations, the remaining frequency components in Figure~\ref{fig:weak}(b) have a strong temperature dependence and are only visible at low temperatures. They are quantum oscillations arising from different Fermi surfaces. The frequency of quantum oscillations from a closed Fermi surfaces is related to the area of the Fermi surface through
\begin{equation}\label{eq:onsagerRule}
    f(A)=\frac{\hbar A}{2\pi e}
\end{equation}
where $A$ is the area of the Fermi surface. The area of the reconstructed Fermi surfaces can be calculated geometrically using the original circular Fermi surface, whose radius is $k_F=\sqrt{2\pi n_s}$ (where $n_s$ is the electron density). From the experimental data, we identify three different Fermi surfaces $A_0$, $A_1$ and $2A_0-A_1$ as depicted in Figure~\ref{fig:Fermi}(b). The calculated frequencies of the quantum oscillations from these Fermi surfaces are plotted as red, blue and green dashed lines in Figure~\ref{fig:weak}(b) and are in excellent agreement with the experiment. 

The observation of reconstructed Fermi surfaces $A_1$ and $2A_0-A_1$ is strongly related to the process of magnetic breakdown.
The probability of electrons completing loops around different Fermi surfaces depends on their probability of tunnelling through the energy gap at the anticrossing between different Brillouin zones (Figure~\ref{fig:device}(d)). For a weak magnetic field $B \to 0$, the tunnelling probability is proportional to the size of the gap, which is directly related to the strength of the electrostatic modulation from the periodic lattice. If the modulation is not too strong, the gap is small and the tunnelling probability is high. Therefore, only Fermi surfaces $A_1$ and $2A_0-A_1$ with two reflections, the minimum number of reflections required to define a reconstructed Fermi surface, are observed experimentally. The higher visibility of $A_1$ compared to $2A_0-A_1$ in Figure~\ref{fig:weak}(b) can be understood by the shorter real space trajectory of $A_1$ than $2A_0-A_1$. This means electrons are more likely to complete orbits defined by $A_1$ rather than $2A_0-A_1$, which makes $A_1$ the main reconstructed Fermi surface in the system.

The process of magnetic breakdown also explains the braiding pattern of $\Delta R_{xx}$ in Figure~\ref{fig:weak}(a). At large magnetic fields, tunneling probability is high and the original Fermi surface $A_0$ is fully restored due to magnetic breakdown. Therefore, at magnetic fields larger than $0.3\ \mathrm{T}$, only the Landau fan of states associated with $A_0$ are visible. At lower magnetic fields, $B<0.3\ \mathrm{T}$, magnetic breakdown becomes weak and electrons start to follow the reconstructed Fermi surface $A_1$. Unlike $A_0$, the Landau fan of $A_1$ emanates from a finite electron density of $n=1.6\times 10^{10}\ \mathrm{cm^{-2}}$. This is the electron density at which $k_F$, the radius of $A_0$, is equal to the reciprocal lattice vector $\bm{g}$ and thus the density where $A_1$ begins to form. The co-existence of the $A_0$ and $A_1$ Landau fans at intermediate fields results in the braiding pattern in $\Delta R_{xx}$ as illustrated by the calculated fan diagram in Figure~\ref{fig:Fermi}(c).

There are two additional frequency components (indicated by black arrows in Figure~\ref{fig:weak}(b)) which  similarly to the $A_{1}$ and $A_{0}$ lines, disappear at $T = 1.5 \ \mathrm{K}$. Their frequencies correspond to fractional combinations of the $A_{1}$ and $A_{0}$ lines, i.e. $f_{1} = f( (A_{0} + A_{1}) / 2 )$ and $f_{2} = f( ( 3 A_{0} - A_{1} ) / 2 ) $. These frequencies can also be interpreted as averages: $f_{1}$ is the average of $f_{A_{0}}$ and $f_{A_{1}}$ while $f_{2}$ is the average of $f_{A_{0}}$ and $f_{2A_{0} - A_{1}}$. Neither $f_{1}$ nor $f_{2}$ can be understood as arising from closed Fermi surface trajectories with area $(A_{0} + A_{1}) / 2$ or $( 3 A_{0} - A_{1} ) / 2$ since no such trajectories exists in the triangular lattice Fermi surface. Lines with these frequencies have not been observed before in 1D~\cite{Deutschmann01} or 2D lattices~\cite{Albrecht99} and are not currently understood.


\begin{figure}[h!]
\centering
\includegraphics[width=\textwidth]{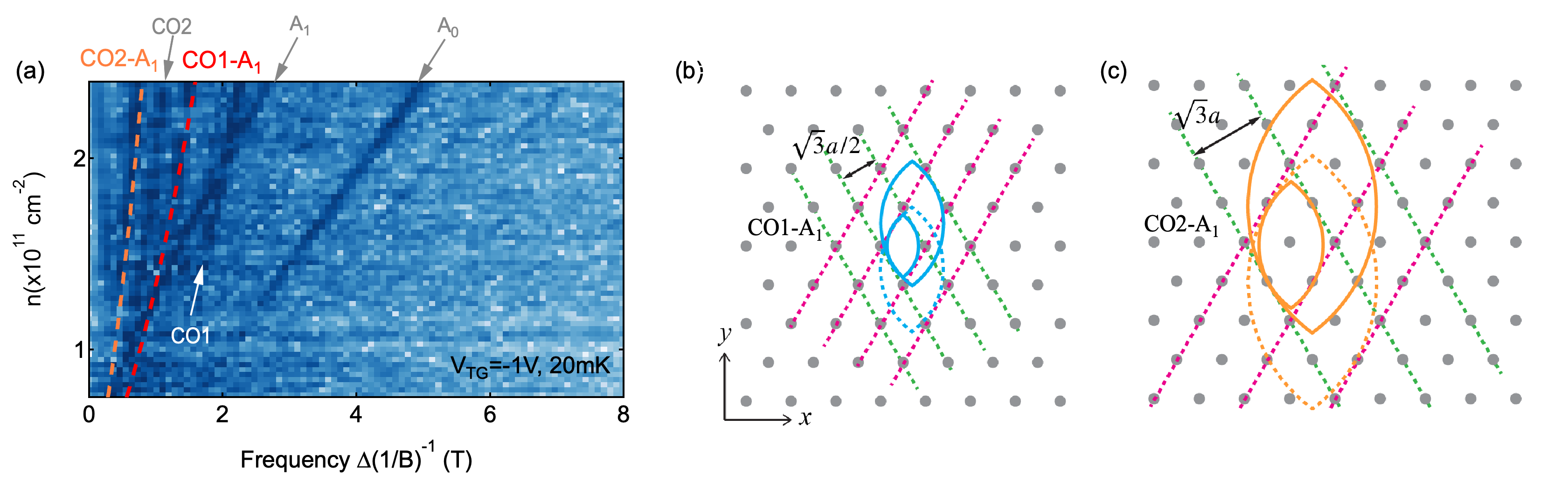}
\caption{(a) Fourier transforms of the magnetoresistance $\Delta R_{xx}$ as a function of $1/B$ for $\mathrm{V_{TG}=-1\ V}$ at $\mathrm{T=20\ mK}$. The dashed lines are the theoretically calculated frequencies of the commensurability oscillations of $A_1$ for single-lattice periodic (red, CO1-A$_1$) and double-lattice periodic (orange, CO2-A$_1$) commensurate conditions respectively. (b) Schematic depicting the real space commensurate conditions of electrons from the reconstructed Fermi surface $A_1$ scattering from a 2D triangular lattice. Here $A_1$ is the reconstructed Fermi surface caused by the horizontal lattice planes (rotated by $90^{\circ}$ due to the translation from reciprocal to the real space). Green and pink dashed lines indicate the two lattice planes which are both $60^{\circ}$ off the horizontal plane. The solid blue ovals show the electron trajectories, scattering from the green lattice planes for different magnetic fields, causing the CO1-A$_1$ commensurability oscillations. The dashed blue oval shows $A_1$ electron scattering from the pink lattice planes, $120^{\circ}$ to the green lattice planes, which also results in commensurability oscillations with the same frequency as from the green lattice planes. (c) Due to the symmetry of the triangular lattice, electrons from the $A_1$ Fermi surface can also scatter from every second lattice plane, resulting in CO2-A$_1$ commensurability oscillations.}
\label{fig:Strong}
\end{figure}

Finally we examine how the system behaves when the modulation strength is increased. Increasing the negative bias on the top-gate strengthens the modulation potential. Figure~\ref{fig:Strong}(a) shows the Fourier transform of background removed magnetoresistance $\Delta R_{xx}$ at $T=20\ \mathrm{mK}$ as a function of $1/B$ for $\mathrm{V_{TG}=-1\ V}$.  Comparing Figure~\ref{fig:weak}(b) at $\mathrm{V_{TG}=0.5\ V}$ to Figure~\ref{fig:Strong}(a) at $\mathrm{V_{TG}=-1\ V}$,  two extra frequency components are observed and highlighted by the dashed lines labelled CO1-A$_1$ and CO2-A$_1$. We trace these frequencies to the commensurability oscillations of the reconstructed Fermi surface $A_1$ (see supplementary section~\ref{sup:temp} for more details). As shown in Figure~\ref{fig:Fermi}(b), the formation of $A_1$ only requires lattice planes in one dimension. Therefore, for a 2D triangular lattice, electrons from Fermi surface $A_1$ can scatter from the other two sets of lattice planes, at $\pm 60^{\circ}$ giving rise to commensurability oscillations of the reconstructed Fermi surface as shown in Figure~\ref{fig:Strong}(b). The calculated frequency of this commensurate condition of $A_1$ (see supplementary section~\ref{supp:frequency}) is plotted as the red dashed line in Figure~\ref{fig:Strong}(a) labelled CO1-A$_1$. Unique to the triangular lattice, every second lattice plane can give rise to another set of commensurability oscillations of the reconstructed Fermi surface as shown in Figure~\ref{fig:Strong}(c). This double spacing condition thus produces commensurability oscillations at a frequency exactly half of CO1-A$_1$, which is plotted as the orange dashed line labelled CO2-A$_1$ in Figure~\ref{fig:Strong}(a). The calculated frequencies of both CO1-A$_1$ and CO2-A$_1$ are in excellent agreement with the experiment, again with no fitting parameters.




In conclusion, we observe oscillations in magnetoresistance that are unambiguously linked to the artificial bandstructure due to a 2D lateral superlattice. Furthermore, as we increase the modulation potential we observe, for the first time, the classical commensurability oscillations of the reconstructed Fermi surface $A_1$ at elevated temperatures. Our work demonstrates a path to bandstructure engineering  in conventional semiconductor systems, which could open new opportunities towards studying the physics of artificial graphene,  strongly correlated systems, superconductivity and magnetism, as well as artificial 2D topological insulators~\cite{Park09,Sushkov13,Du21,Polini13}.

\begin{acknowledgement}
 This work was funded by the Australian Research Council Centre of Excellence for Future Low Energy Electronics Technologies (CE170100039) and EP/R029075/1 Non-Ergodic Quantum Manipulation, UK. Device fabrication was partially carried out at the Australian National Fabrication Facility (ANFF) at the UNSW node. 
\end{acknowledgement}

\begin{suppinfo}
    
\renewcommand{\thefigure}{S\arabic{figure}}
\setcounter{figure}{0}  

\subsection{Calculation of frequency lines in the Fourier transform of magnetoresistance oscillations }
\label{supp:frequency}
There are two kinds of frequency lines that we compute: there are the quantum oscillations which come from Onsager's formula, and there are commensurability oscillations which come from a generalisation of the standard commensurability condition. Both of these kinds of frequencies depend on the geometry of the Fermi surface: the quantum frequencies are proportional to the area of the orbit, and the commensurability frequencies are proportional to the width of the orbit in a particular direction. In all of these calculations it is sufficient to compute the area in the limit that the modulation strength goes to zero. For the quantum oscillations, the areas of $A_{0}$ and $A_{1}$ are given by

\begin{align*}
    A_{0} &= \pi k_{F}^{2} \\
    A_{1} &= \pi k_{F}^{2} - g \sqrt{k_{F}^{2} - g^{2} / 4}
             - 2 k_{F}^{2} \arctan\left( \frac{g / 2}{\sqrt{k_{F}^{2} - g^{2} / 4}} \right)
\end{align*}

The equation for $A_{1}$ is the area of intersection between two circles of radius $k_{F}$ and separation $g$. The frequencies $f(A_{0})$, $f(A_{1})$, and $f(2 A_{0} - A_{1})$, plotted in Figure \ref{fig:weak}, can then be obtained using equation Eqn. \ref{eq:onsagerRule}.

Commensurability frequencies come from comparing the width of an orbit in real space to a certain set of lattice planes. For the standard commensurability oscillations (CO1 in Fig. \ref{fig:Fermi}a) we are comparing the width, $\tilde{w} = 2 \hbar k_{F} / e B$, of $A_{0}$ to the spacing between lattice planes, $L = \sqrt{3} a / 2$, to obtain a frequency

\begin{align*}
    f = \frac{\hbar w}{eL}
\end{align*}

Where $w$ is the width of the orbit in $k$-space. This is equivalent to equation Eqn. \ref{eq:co2}. The frequency of CO2 (see Fig. \ref{fig:Fermi}a) is then obtained by doubling the lattice plane spacing to $L = \sqrt{3} a$, which halves the frequency. More general frequencies are obtained by comparing the width of $A_{1}$ in real space to the set of lattice planes that make a 30 degree angle with the major axis of $A_{1}$. This was depicted in Fig. \ref{fig:Strong}b. The real space width, $\tilde{w} = \hbar w / e B$, is the total width of the orbit along a line perpendicular to the lattice planes, that is, a line making a 60 degree angle to the major axis of $A_{1}$. In $k$-space this width turns out to be equal to

\begin{align*}
    w = 2 k_{F} - \sqrt{3} g / 2
\end{align*}

And, with $L = \sqrt{3} a / 2$, we obtain a frequency

\begin{align*}
    f = \frac{4 \hbar k_{F} - \sqrt{3} \hbar g}{e \sqrt{3} a}
\end{align*}

which is the frequency labelled CO1-A$_1$ in Fig. \ref{fig:Strong}a. Again, to obtain the half-frequency line (CO2-A$_1$ in Fig. \ref{fig:Strong}a) we double the value of $L$ (Fig. \ref{fig:Strong}c). We thus have a set of frequencies that are functions of the Fermi momentum $k_{F}$. To obtain frequency as a function of electron density, $n$, we use $k_{F} = \sqrt{2 \pi n}$.

\subsection{Background removal of $R_{xx}$}
\label{sup:background}
In the main text, background removed longitudinal resistance $\Delta R_{xx}$ is plotted instead of $R_{xx}$ to emphasize the resistance oscillations. $\Delta R_{xx}$ is also used in the Fourier analysis. In Figure~\ref{fig:supp background} we show the original $R_{xx}$, the smoothed background $R_b$ and the resultant $\Delta R_{xx} = R_{xx} - R_b$ for two different densities.

\begin{figure}[h!]
\centering
\includegraphics[width=\textwidth]{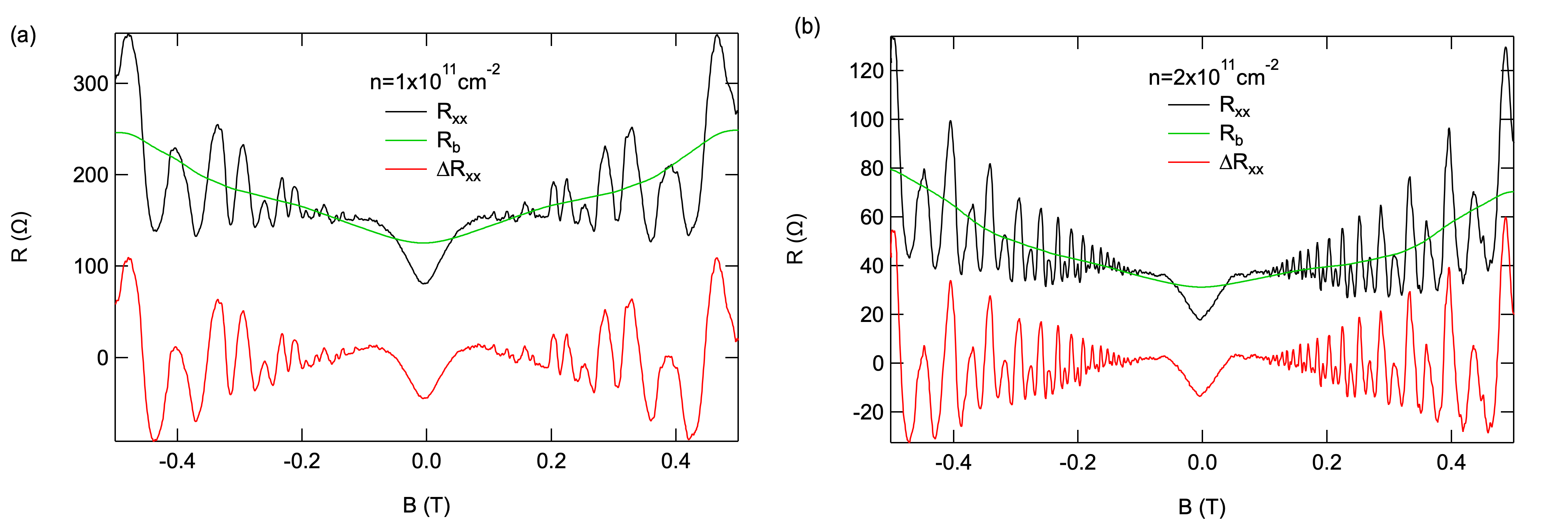}
\caption{Measured longitudinal resistance $R_{xx}$, the smoothed background $R_b$ and background removed longitudinal resistance $\Delta R_{xx}$ for two densities (a) $1\times10^{11}cm^{-2}$ and (b) $2\times10^{11}cm^{-2}$ respectively.}
\label{fig:supp background}
\end{figure}

\subsection{The effect of temperature on COs of the $A_{1}$ Fermi surface }
\label{sup:temp}

To identify the origin of new oscillations when the modulation strength is increased, magnetoresistance measurements and subsequent Fourier analysis are performed at an elevated temperature of $1.5\ \mathrm{K}$. As shown in Figure~\ref{fig:supp temp}(a), at the same modulation strength $\mathrm{V_{TG}=-1\ V}$ as in Figure~\ref{fig:Strong}(a), new frequency components (CO1-A$_1$ and CO2-A$_1$) become very weak at $1.5\ \mathrm{K}$. However, if the modulation strength is increased further  to $\mathrm{V_{TG}=-2\ V}$, CO1-A$_1$ and CO2-A$_1$ persist to higher temperatures as can be seen in Figure~\ref{fig:supp temp}(b). The existence of these frequency components at $1.5\ \mathrm{K}$ strongly suggests that they originate from classical commensurability oscillations, although a stronger modulation potential is required to observe them. This is because thermal excitation increases the probability of electrons tunnelling through the energy gap between different orbits and thus prohibits the formation of the reconstructed Fermi surface $A_1$. Therefore, a stronger modulation potential (i.e. larger energy gap) is required to observe the commensurability oscillations of $A_1$ at elevated temperatures.

\begin{figure}[h!]
\centering
\includegraphics[width=\textwidth]{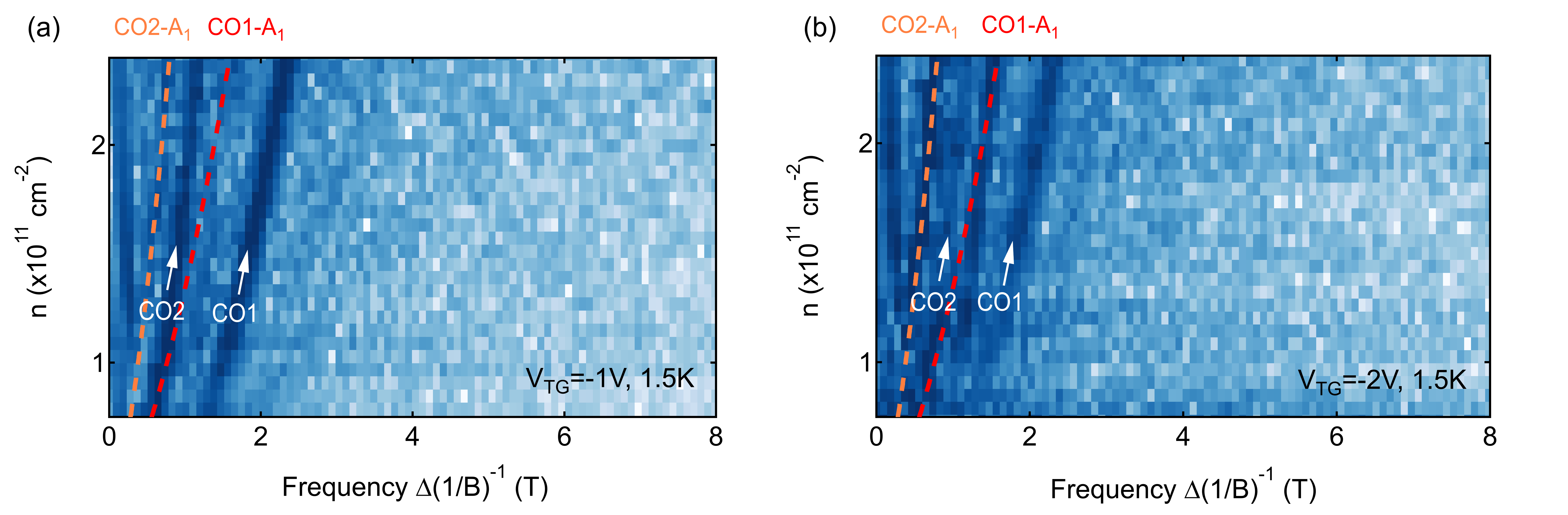}
\caption{(a) Fourier transforms of the magnetoresistance $\Delta R_{xx}$ as a function of $1/B$ for (a) $\mathrm{V_{TG}=-1\ V}$ at $\mathrm{T=1.5\ K}$ and (b) $\mathrm{V_{TG}=-2\ V}$ at $\mathrm{T=1.5\ K}$. The dashed lines are the theoretically calculated frequencies of the commensurability oscillations of $A_1$ for single spacing (red, CO1-A$_1$) and double spacing (orange, CO2-A$_1$) commensurate conditions respectively.}
\label{fig:supp temp}
\end{figure}

\end{suppinfo}

\bibliography{references}

\providecommand{\latin}[1]{#1}
\makeatletter
\providecommand{\doi}
  {\begingroup\let\do\@makeother\dospecials
  \catcode`\{=1 \catcode`\}=2 \doi@aux}
\providecommand{\doi@aux}[1]{\endgroup\texttt{#1}}
\makeatother
\providecommand*\mcitethebibliography{\thebibliography}
\csname @ifundefined\endcsname{endmcitethebibliography}
  {\let\endmcitethebibliography\endthebibliography}{}
\begin{mcitethebibliography}{26}
\providecommand*\natexlab[1]{#1}
\providecommand*\mciteSetBstSublistMode[1]{}
\providecommand*\mciteSetBstMaxWidthForm[2]{}
\providecommand*\mciteBstWouldAddEndPuncttrue
  {\def\EndOfBibitem{\unskip.}}
\providecommand*\mciteBstWouldAddEndPunctfalse
  {\let\EndOfBibitem\relax}
\providecommand*\mciteSetBstMidEndSepPunct[3]{}
\providecommand*\mciteSetBstSublistLabelBeginEnd[3]{}
\providecommand*\EndOfBibitem{}
\mciteSetBstSublistMode{f}
\mciteSetBstMaxWidthForm{subitem}{(\alph{mcitesubitemcount})}
\mciteSetBstSublistLabelBeginEnd
  {\mcitemaxwidthsubitemform\space}
  {\relax}
  {\relax}

\bibitem[Dean \latin{et~al.}(2013)Dean, Wang, Maher, Forsythe, Ghahari, Gao,
  Katoch, Ishigami, Moon, Koshino, Taniguchi, Watanabe, Shepard, Hone, and
  Kim]{Dean13}
Dean,~C.; Wang,~L.; Maher,~P.; Forsythe,~C.; Ghahari,~F.; Gao,~Y.; Katoch,~J.;
  Ishigami,~M.; Moon,~P.; Koshino,~M.; Taniguchi,~T.; Watanabe,~K.;
  Shepard,~K.; Hone,~J.; Kim,~P. Hofstadter’s butterfly and the fractal
  quantum Hall effect in moire´ superlattices. \emph{Nature} \textbf{2013},
  \emph{497}, 598--602\relax
\mciteBstWouldAddEndPuncttrue
\mciteSetBstMidEndSepPunct{\mcitedefaultmidpunct}
{\mcitedefaultendpunct}{\mcitedefaultseppunct}\relax
\EndOfBibitem
\bibitem[Cao \latin{et~al.}(2018)Cao, Fatemi, Demir, Fang, Tomarken, Luo,
  Sanchez-Yamagishi, Watanabe, Taniguchi, Kaxiras, Ashoori, and
  Jarillo-Herrero]{Cao18}
Cao,~Y.; Fatemi,~V.; Demir,~A.; Fang,~S.; Tomarken,~S.~L.; Luo,~J.~Y.;
  Sanchez-Yamagishi,~J.~D.; Watanabe,~K.; Taniguchi,~T.; Kaxiras,~E.;
  Ashoori,~R.~C.; Jarillo-Herrero,~P. Correlated insulator behaviour at
  half-filling in magic-angle graphene superlattices. \emph{Nature}
  \textbf{2018}, \emph{556}, 80--84\relax
\mciteBstWouldAddEndPuncttrue
\mciteSetBstMidEndSepPunct{\mcitedefaultmidpunct}
{\mcitedefaultendpunct}{\mcitedefaultseppunct}\relax
\EndOfBibitem
\bibitem[Cao \latin{et~al.}(2018)Cao, Fatemi, Fang, Watanabe, Taniguchi,
  Kaxiras, and Jarillo-Herrero]{Cao18_2}
Cao,~Y.; Fatemi,~V.; Fang,~S.; Watanabe,~K.; Taniguchi,~T.; Kaxiras,~E.;
  Jarillo-Herrero,~P. Unconventional superconductivity in magic-angle graphene
  superlattices. \emph{Nature} \textbf{2018}, \emph{556}, 43--50\relax
\mciteBstWouldAddEndPuncttrue
\mciteSetBstMidEndSepPunct{\mcitedefaultmidpunct}
{\mcitedefaultendpunct}{\mcitedefaultseppunct}\relax
\EndOfBibitem
\bibitem[Bistritzer and MacDonald(2011)Bistritzer, and MacDonald]{Bistritzer11}
Bistritzer,~R.; MacDonald,~A. Moiré bands in twisted double-layer graphene.
  \emph{Proc. Natl Acad. Sci.} \textbf{2011}, \emph{108}, 12233–12237\relax
\mciteBstWouldAddEndPuncttrue
\mciteSetBstMidEndSepPunct{\mcitedefaultmidpunct}
{\mcitedefaultendpunct}{\mcitedefaultseppunct}\relax
\EndOfBibitem
\bibitem[Liu \latin{et~al.}(2020)Liu, Hao, Khalaf, Lee, Ronen, Yoo,
  Haei~Najafabadi, Watanabe, Taniguchi, Vishwanath, and Kim]{Liu20}
Liu,~X.; Hao,~Z.; Khalaf,~E.; Lee,~J.~Y.; Ronen,~Y.; Yoo,~H.;
  Haei~Najafabadi,~D.; Watanabe,~K.; Taniguchi,~T.; Vishwanath,~A.; Kim,~P.
  Tunable spin-polarized correlated states in twisted double bilayer graphene.
  \emph{Nature} \textbf{2020}, \emph{583}, 221--225\relax
\mciteBstWouldAddEndPuncttrue
\mciteSetBstMidEndSepPunct{\mcitedefaultmidpunct}
{\mcitedefaultendpunct}{\mcitedefaultseppunct}\relax
\EndOfBibitem
\bibitem[Wang \latin{et~al.}(2020)Wang, Shih, Ghiotto, Xian, Rhodes, Tan,
  Claassen, Kennes, Bai, Kim, Watanabe, Taniguchi, Zhu, Hone, Rubio, Pasupathy,
  and Dean]{Wang20_2}
Wang,~L. \latin{et~al.}  Correlated electronic phases in twisted bilayer
  transition metal dichalcogenides. \emph{Nature Materials} \textbf{2020},
  \emph{19}, 861--866\relax
\mciteBstWouldAddEndPuncttrue
\mciteSetBstMidEndSepPunct{\mcitedefaultmidpunct}
{\mcitedefaultendpunct}{\mcitedefaultseppunct}\relax
\EndOfBibitem
\bibitem[Lu \latin{et~al.}(2019)Lu, Stepanov, Yang, Xie, Aamir, Das, Urgell,
  Watanabe, Taniguchi, Zhang, Bachtold, MacDonald, and Efetov]{Lu19}
Lu,~X.; Stepanov,~P.; Yang,~W.; Xie,~M.; Aamir,~M.~A.; Das,~I.; Urgell,~C.;
  Watanabe,~K.; Taniguchi,~T.; Zhang,~G.; Bachtold,~A.; MacDonald,~A.~H.;
  Efetov,~D.~K. Superconductors, orbital magnets and correlated states in
  magic-angle bilayer graphene. \emph{Nature} \textbf{2019}, \emph{574},
  653--657\relax
\mciteBstWouldAddEndPuncttrue
\mciteSetBstMidEndSepPunct{\mcitedefaultmidpunct}
{\mcitedefaultendpunct}{\mcitedefaultseppunct}\relax
\EndOfBibitem
\bibitem[Uri \latin{et~al.}(2020)Uri, Grover, Cao, Crosse, Bagani,
  Rodan-Legrain, Myasoedov, Watanabe, Taniguchi, Moon, Koshino,
  Jarillo-Herrero, and Zeldov]{Uri20}
Uri,~A.; Grover,~S.; Cao,~Y.; Crosse,~J.~A.; Bagani,~K.; Rodan-Legrain,~D.;
  Myasoedov,~Y.; Watanabe,~K.; Taniguchi,~T.; Moon,~P.; Koshino,~M.;
  Jarillo-Herrero,~P.; Zeldov,~E. Mapping the twist-angle disorder and Landau
  levels in magic-angle graphene. \emph{Nature} \textbf{2020}, \emph{581},
  47--52\relax
\mciteBstWouldAddEndPuncttrue
\mciteSetBstMidEndSepPunct{\mcitedefaultmidpunct}
{\mcitedefaultendpunct}{\mcitedefaultseppunct}\relax
\EndOfBibitem
\bibitem[Forsythe \latin{et~al.}(2018)Forsythe, Zhou, Watanabe, Taniguchi,
  Pasupathy, Moon, Koshino, Kim, and Dean]{Forsythe18}
Forsythe,~C.; Zhou,~X.; Watanabe,~K.; Taniguchi,~T.; Pasupathy,~A.; Moon,~P.;
  Koshino,~M.; Kim,~P.; Dean,~C.~R. Band structure engineering of 2D materials
  using patterned dielectric superlattices. \emph{Nature Nanotechnology}
  \textbf{2018}, \emph{13}, 566--571\relax
\mciteBstWouldAddEndPuncttrue
\mciteSetBstMidEndSepPunct{\mcitedefaultmidpunct}
{\mcitedefaultendpunct}{\mcitedefaultseppunct}\relax
\EndOfBibitem
\bibitem[Jessen \latin{et~al.}(2019)Jessen, Gammelgaard, Thomsen, Mackenzie,
  Thomsen, Caridad, Duegaard, Watanabe, Taniguchi, Booth, Pedersen, Jauho, and
  B{\o}ggild]{Jessen19}
Jessen,~B.~S.; Gammelgaard,~L.; Thomsen,~M.~R.; Mackenzie,~D. M.~A.;
  Thomsen,~J.~D.; Caridad,~J.; Duegaard,~E.; Watanabe,~K.; Taniguchi,~T.;
  Booth,~T.~J.; Pedersen,~T.~G.; Jauho,~A.-P.; B{\o}ggild,~P. Lithographic band
  structure engineering of graphene. \emph{Nature Nanotechnology}
  \textbf{2019}, \emph{14}, 340--346\relax
\mciteBstWouldAddEndPuncttrue
\mciteSetBstMidEndSepPunct{\mcitedefaultmidpunct}
{\mcitedefaultendpunct}{\mcitedefaultseppunct}\relax
\EndOfBibitem
\bibitem[Huber \latin{et~al.}(2022)Huber, Steffen, Drienovsky, Sandner,
  Watanabe, Taniguchi, Pfannkuche, Weiss, and Eroms]{Huber22}
Huber,~R.; Steffen,~M.~N.; Drienovsky,~M.; Sandner,~A.; Watanabe,~K.;
  Taniguchi,~T.; Pfannkuche,~D.; Weiss,~D.; Eroms,~J. Band conductivity
  oscillations in a gate-tunable graphene superlattice. \emph{Nature
  Communications} \textbf{2022}, \emph{13}, 2856\relax
\mciteBstWouldAddEndPuncttrue
\mciteSetBstMidEndSepPunct{\mcitedefaultmidpunct}
{\mcitedefaultendpunct}{\mcitedefaultseppunct}\relax
\EndOfBibitem
\bibitem[Gerhardts \latin{et~al.}(1989)Gerhardts, Weiss, and
  Klitzing]{Gerhardts89}
Gerhardts,~R.~R.; Weiss,~D.; Klitzing,~K.~v. Novel magnetoresistance
  oscillations in a periodically modulated two-dimensional electron gas.
  \emph{Phys. Rev. Lett.} \textbf{1989}, \emph{62}, 1173--1176\relax
\mciteBstWouldAddEndPuncttrue
\mciteSetBstMidEndSepPunct{\mcitedefaultmidpunct}
{\mcitedefaultendpunct}{\mcitedefaultseppunct}\relax
\EndOfBibitem
\bibitem[Winkler \latin{et~al.}(1989)Winkler, Kotthaus, and Ploog]{Winkler89}
Winkler,~R.~W.; Kotthaus,~J.~P.; Ploog,~K. Landau band conductivity in a
  two-dimensional electron system modulated by an artificial one-dimensional
  superlattice potential. \emph{Phys. Rev. Lett.} \textbf{1989}, \emph{62},
  1177--1180\relax
\mciteBstWouldAddEndPuncttrue
\mciteSetBstMidEndSepPunct{\mcitedefaultmidpunct}
{\mcitedefaultendpunct}{\mcitedefaultseppunct}\relax
\EndOfBibitem
\bibitem[Albrecht \latin{et~al.}(1999)Albrecht, Smet, Weiss, von Klitzing,
  Hennig, Langenbuch, Suhrke, R\"ossler, Umansky, and Schweizer]{Albrecht99}
Albrecht,~C.; Smet,~J.~H.; Weiss,~D.; von Klitzing,~K.; Hennig,~R.;
  Langenbuch,~M.; Suhrke,~M.; R\"ossler,~U.; Umansky,~V.; Schweizer,~H.
  Fermiology of Two-Dimensional Lateral Superlattices. \emph{Phys. Rev. Lett.}
  \textbf{1999}, \emph{83}, 2234--2237\relax
\mciteBstWouldAddEndPuncttrue
\mciteSetBstMidEndSepPunct{\mcitedefaultmidpunct}
{\mcitedefaultendpunct}{\mcitedefaultseppunct}\relax
\EndOfBibitem
\bibitem[Deutschmann \latin{et~al.}(2001)Deutschmann, Wegscheider, Rother,
  Bichler, Abstreiter, Albrecht, and Smet]{Deutschmann01}
Deutschmann,~R.~A.; Wegscheider,~W.; Rother,~M.; Bichler,~M.; Abstreiter,~G.;
  Albrecht,~C.; Smet,~J.~H. Quantum Interference in Artificial Band Structures.
  \emph{Phys. Rev. Lett.} \textbf{2001}, \emph{86}, 1857--1860\relax
\mciteBstWouldAddEndPuncttrue
\mciteSetBstMidEndSepPunct{\mcitedefaultmidpunct}
{\mcitedefaultendpunct}{\mcitedefaultseppunct}\relax
\EndOfBibitem
\bibitem[Geisler \latin{et~al.}(2004)Geisler, Smet, Umansky, von Klitzing,
  Naundorf, Ketzmerick, and Schweizer]{Geisler04}
Geisler,~M.~C.; Smet,~J.~H.; Umansky,~V.; von Klitzing,~K.; Naundorf,~B.;
  Ketzmerick,~R.; Schweizer,~H. Detection of a Landau Band-Coupling-Induced
  Rearrangement of the Hofstadter Butterfly. \emph{Phys. Rev. Lett.}
  \textbf{2004}, \emph{92}, 256801\relax
\mciteBstWouldAddEndPuncttrue
\mciteSetBstMidEndSepPunct{\mcitedefaultmidpunct}
{\mcitedefaultendpunct}{\mcitedefaultseppunct}\relax
\EndOfBibitem
\bibitem[Srinivasan \latin{et~al.}(2020)Srinivasan, Farrer, Ritchie, and
  Hamilton]{Ashwin20}
Srinivasan,~A.; Farrer,~I.; Ritchie,~D.~A.; Hamilton,~A.~R. Improving
  reproducibility of quantum devices with completely undoped architectures.
  \emph{Applied Physics Letters} \textbf{2020}, \emph{117}, 183101\relax
\mciteBstWouldAddEndPuncttrue
\mciteSetBstMidEndSepPunct{\mcitedefaultmidpunct}
{\mcitedefaultendpunct}{\mcitedefaultseppunct}\relax
\EndOfBibitem
\bibitem[Tkachenko \latin{et~al.}(2015)Tkachenko, Tkachenko, Terekhov, and
  Sushkov]{Tkachenko15}
Tkachenko,~O.~A.; Tkachenko,~V.~A.; Terekhov,~I.~S.; Sushkov,~O.~P. Effects of
  Coulomb screening and disorder on an artificial graphene based on
  nanopatterned semiconductor. \emph{2D Materials} \textbf{2015}, \emph{2},
  014010\relax
\mciteBstWouldAddEndPuncttrue
\mciteSetBstMidEndSepPunct{\mcitedefaultmidpunct}
{\mcitedefaultendpunct}{\mcitedefaultseppunct}\relax
\EndOfBibitem
\bibitem[Wang \latin{et~al.}(2020)Wang, Reuter, Wieck, Hamilton, and
  Klochan]{Wang20}
Wang,~D.~Q.; Reuter,~D.; Wieck,~A.~D.; Hamilton,~A.~R.; Klochan,~O.
  Two-dimensional lateral surface superlattices in GaAs heterostructures with
  independent control of carrier density and modulation potential.
  \emph{Applied Physics Letters} \textbf{2020}, \emph{117}, 032102\relax
\mciteBstWouldAddEndPuncttrue
\mciteSetBstMidEndSepPunct{\mcitedefaultmidpunct}
{\mcitedefaultendpunct}{\mcitedefaultseppunct}\relax
\EndOfBibitem
\bibitem[Steda and MacDonald(1990)Steda, and MacDonald]{Steda90}
Steda,~P.; MacDonald,~A.~H. Magnetic breakdown and magnetoresistance
  oscillations in a periodically modulated two-dimensional electron gas.
  \emph{Phys. Rev. B} \textbf{1990}, \emph{41}, 11892--11898\relax
\mciteBstWouldAddEndPuncttrue
\mciteSetBstMidEndSepPunct{\mcitedefaultmidpunct}
{\mcitedefaultendpunct}{\mcitedefaultseppunct}\relax
\EndOfBibitem
\bibitem[Beenakker(1989)]{Beenakker89}
Beenakker,~C. W.~J. Guiding-center-drift resonance in a periodically modulated
  two-dimensional electron gas. \emph{Phys. Rev. Lett.} \textbf{1989},
  \emph{62}, 2020--2023\relax
\mciteBstWouldAddEndPuncttrue
\mciteSetBstMidEndSepPunct{\mcitedefaultmidpunct}
{\mcitedefaultendpunct}{\mcitedefaultseppunct}\relax
\EndOfBibitem
\bibitem[Park and Louie(2009)Park, and Louie]{Park09}
Park,~C.-H.; Louie,~S. Making Massless Dirac Fermions from a Patterned
  Two-Dimensional Electron Gas. \emph{Nano Letters} \textbf{2009}, \emph{9},
  1793--1797\relax
\mciteBstWouldAddEndPuncttrue
\mciteSetBstMidEndSepPunct{\mcitedefaultmidpunct}
{\mcitedefaultendpunct}{\mcitedefaultseppunct}\relax
\EndOfBibitem
\bibitem[Sushkov and Castro~Neto(2013)Sushkov, and Castro~Neto]{Sushkov13}
Sushkov,~O.~P.; Castro~Neto,~A.~H. Topological Insulating States in Laterally
  Patterned Ordinary Semiconductors. \emph{Phys. Rev. Lett.} \textbf{2013},
  \emph{110}, 186601\relax
\mciteBstWouldAddEndPuncttrue
\mciteSetBstMidEndSepPunct{\mcitedefaultmidpunct}
{\mcitedefaultendpunct}{\mcitedefaultseppunct}\relax
\EndOfBibitem
\bibitem[Du \latin{et~al.}(2021)Du, Liu, Wind, Pellegrini, West, Fallahi,
  Pfeiffer, Manfra, and Pinczuk]{Du21}
Du,~L.; Liu,~Z.; Wind,~S.; Pellegrini,~V.; West,~K.; Fallahi,~S.; Pfeiffer,~L.;
  Manfra,~M.; Pinczuk,~A. Observation of Flat Bands in Gated Semiconductor
  Artificial Graphene. \emph{Physical Review Letters} \textbf{2021},
  \emph{126}, 106402\relax
\mciteBstWouldAddEndPuncttrue
\mciteSetBstMidEndSepPunct{\mcitedefaultmidpunct}
{\mcitedefaultendpunct}{\mcitedefaultseppunct}\relax
\EndOfBibitem
\bibitem[Polini \latin{et~al.}(2013)Polini, Guinea, Lewenstein, Manoharan, and
  Pellegrini]{Polini13}
Polini,~M.; Guinea,~F.; Lewenstein,~M.; Manoharan,~H.~C.; Pellegrini,~V.
  Artificial honeycomb lattices for electrons, atoms and photons. \emph{Nature
  Nanotechnology} \textbf{2013}, \emph{8}, 625\relax
\mciteBstWouldAddEndPuncttrue
\mciteSetBstMidEndSepPunct{\mcitedefaultmidpunct}
{\mcitedefaultendpunct}{\mcitedefaultseppunct}\relax
\EndOfBibitem
\end{mcitethebibliography}

\end{document}